\documentstyle[12pt]{article}
\newcommand{\be}{\begin{equation}}
\newcommand{\ee}{\end{equation}}
\newcommand{\ba}{\begin{array}}
\newcommand{\ea}{\end{array}}
\newcommand{\bc}{\begin{center}}
\newcommand{\ec}{\end{center}}
\newcommand{\disregard}[1]{{}}
\newcommand{\ti}{\tilde}
\newcommand{\al}{\alpha}
\newcommand{\la}{\lambda}
\newcommand{\am}{\langle\al\rangle}
\newcommand{\op}{b}
\newcommand{\sco}{f}
\newcommand{\zoz}{z}
\newcommand{\mm}{\mu}
\newcommand{\nn}{\nu}
\newcommand{\Om}{{\cal F}}
\newcommand{\ener}{{\cal E}}
\newcommand{\dz}{\delta\zoz}

\newcommand{\xij}{\langle {\bar x}_{i}{\bar x}_{j}\rangle}
\newcommand{\pij}{\langle {\bar p}_{i}{\bar p}_{j}\rangle}
\newcommand{\xipj}{
\langle {\bar x}_{i}{\bar p}_{j}+{\bar p}_{j}{\bar x}_{i}\rangle
}

\newcommand{\uu}{u}
\newcommand{\ww}{w}
\newcommand{\GG}{G}
\newcommand{\FF}{F}
\newcommand{\ds}{\displaystyle}
\newcommand{\demi}{{\ds 1\over\ds 2}}

\newcommand{\rac}{{\ds 1\over\ds\sqrt{2}}}

\newcommand{\Ga}{\Gamma}
\newcommand{\De}{\Delta}
\newcommand{\Lam}{\Lambda}
\newcommand{\ffi}{\varphi}
\newcommand{\py}{\pi}

\newcommand{\prl}[1]{ Phys. Rev. Lett. {\bf #1}}

\newcommand{\apb}[1]{Ann. of Phys. (N.Y.) {\bf #1} }

\title{ On the Poisson Structure of the Time-Dependent 
Mean-Field Equations For Systems of Bosons out of Equilibrium}

\author{
Mohamed Benarous 
\thanks{
Perm. address: University of Chlef, BP 151, 02000 Chlef, Algeria.
}
\\
{\it
Division de Physique Th\'eorique \thanks{Unit\'e 
de Recherche des Universit\'es
Paris XI et Paris VI associ\'ee au C.N.R.S.},
} 
\\
{\it  Institut de Physique Nucl\'eaire,} 
\\
{\it F-91406 , Orsay Cedex, France }
}
\date{\today}
\begin{document}
\maketitle
\begin{abstract}

We analyze the Poisson structure of the time-dependent mean-field equations
for bosons and construct the Lie-Poisson bracket associated to these equations.
The latter follow from the time-dependent variational principle of Balian and
V\'en\'eroni when a gaussian Ansatz is chosen for the density operator.
We perform a stability analysis of both the full and the linearized equations.
We also search for the canonically conjugate variables. In certain cases, the
evolution equations can indeed be cast in a Hamiltonian form.

\end{abstract}



\newpage

\setcounter{section}{0}
\setcounter{equation}{0}
\bc{\section{Introduction}} \ec


Poisson structures have already been used to study various properties
of classical theories such as the Maxwell-Vlasov equations and the BBGKY
hierarchy \cite{marsmor,morrison}. They are useful in quantization methods;
for a review, see \cite{arnold,muku,litt82}.
In the quantum case, besides the abstract quantum group approach, Poisson
structures have found some applications in the many-body problem.
In particular, they have been used to study the stability properties
of important types of approximation such as the static Hartree-Fock
(HF), the time-dependent Hartree-Fock (TDHF) and the random phase 
approximation (RPA), both in the zero temperature \cite{dirac,fbrbtsg} 
and in the thermal case \cite{BV89}. With their help, the connections
between these theories, as well as with classical dynamics, follow in
a very transparent fashion.

The purpose of the present paper is to extend the analysis of
ref.\cite{BV89}, which deals with fermions, to the time evolution
of many-boson systems.
Among the relevant problems that we have in mind, we can quote bose
condensation phenomenon\cite{fettwal,landnozpi,gropi}. In particular,
recent experiments on the Bose-Einstein condensation of dilute atomic
gases in laser traps at ultracold temperatures \cite{gss95} are raising
a wide interest. They involve not only the stu\-dy of equilibrium properties,
but also that of dynamics. In the existing mean-field approaches, this
dynamics deals with the expectation value of a creation or an annihilation
operator, which is the order parameter, and more generally with the
expectation values of product of two such operators. Our purpose is to
analyze the structure of these equations so as to clarify their meaning.

Our paper is organized as follows. We briefly present in section 2.1
the time-dependent variational principle proposed by Balian and
V\'e\-n\'e\-roni (BV)\cite{BV81,BV92} which we use to derive the
approximate evolution equations. We also introduce, as well as some
useful notations, the trial subspaces which we will consider
throughout this work. These ans\"{a}tze lead to the Time-Dependent 
Hartree-Bogoliubov (TDHB) equations. They belong to the class of the
time-dependent mean-field approximations and they are the bosonic
counterpart of the time-dependent Hartree-Fock (TDHF) equations for
fermions. In section 3, we construct the Lie-Poisson bracket associated
with these TDHB equations. This is achieved by extending the analysis of
ref.\cite{BV89} to bose systems. Our formultation includes the possible
existence of bose condensates. We derive in section 4 the corresponding
linearized equations and study their stability in the sense of Lyapunov.
A question that arises naturally in this context is whether canonically
conjugate variables can be found. We perform a canonical transformation
allowing us to rewrite the TDHB equations in a form which, in some
circumstances, is well suited to the discussion of this question.
The last section contains some conclusions.

\setcounter{equation}{0}
\bc{\section{The Bosonic TDHB Equations}}\ec

\noindent
{\bf 2.1. The time-dependent variational principle}

The BV variational principle \cite{BV81,BV92} is devised to evaluate
the expectation value $\hbox{Tr}AD(t_1)$ of a given observable $A$, measured
at a final time $t_1$, when the density operator $D_0$ is known at the initial
time $t_0$. To this aim, one introduces the action-like functional
\be 
\label{e1}
{\cal I}\{{\cal A}(t),{\cal D}(t)\}\equiv
{\rm Tr}\,{\cal A}(t_0) D(t_0)
+\int_{t_0}^{t_1}\,{\rm d}t\,
{\rm Tr}\,{\cal D}(t)
\left(
{\ds {\rm d}{\cal A}(t)\over\ds {\rm d}t}-i[{\cal A}(t),H]
\right)
,
\ee
where the variational quantities are the time-dependent operators 
${\cal D}(t)$ and ${\cal A}(t)$, the latter being subject to the
final boundary condition
\be 
\label{e2}
{\cal A}(t_1)=A .
\ee
The symbol Tr stands for a trace over the complete Fock space (we take
$\hbar$=1). The desired quantity is then provided by the stationary value
of the functional (\ref{e1}), where $H$ is the Hamiltonian of the system.

When the allowed variations $\delta {\cal A}(t)$ and $\delta {\cal D}(t)$
are unrestricted, the stationarity conditions of the BV functional (\ref{e1})
lead to the exact Liouville-von-Neumann equation for ${\cal D}(t)$ and to the
(backward) Heisenberg equation for ${\cal A}(t)$. When restricted trial
ans\"{a}tze are chosen for ${\cal A}(t)$ and ${\cal D}(t)$, the extremization
of (\ref{e1}) yields approximate, and in general coupled, evolution equations
for ${\cal A}(t)$ and ${\cal D}(t)$.

\noindent
{\bf 2.2. Trial spaces and the TDHB equations}

Let us first introduce the trial class of operators, which we shall use
throughout this work for ${\cal D}(t)$ and ${\cal A}(t)$. To this end, we
define the $2n$-component operator $\al$
\be
\label{e3}
\al =\left(
\ba{c}
a^{\phantom{+}}\\
a^{\dag}
\ea\right)
,
\ee
where $a^{\dag}$ and $a$ denote boson creation and annihilation operators
in a given single-particle space of dimension $n$. In terms of the $\al$'s,
the usual commutation relations can be expressed in the compact form
($i,j=1\ldots 2n$)
\be
\label{e4}
[\al_i ,\, \al_j ]=\tau_{ij}\quad ,\quad \tau=i\, \sigma_2
=\pmatrix{\phantom{-}0&1_{n}\cr -1_{n}&\phantom{-}0}
,
\ee
where $1_{n}$ is the $(n\times n)$ unit matrix and $\sigma_2$ the
$(2n\times 2n)$ second Pauli matrix.

With this notation, our trial class for the operator ${\cal D}(t)$, the
exponentials of the linear plus quadratic forms in $\al$, can be written
in the factorized form \cite{BM91,BB69}
\be
\label{e5}
{\cal D}=\exp{(\nu)}
\exp{(\ti{\la}\tau\al)}
\exp{(\demi\ti{\al}\tau S\al)}
,
\ee
where $\nu$ is a c-number, $\la$ a $2n$-component vector and $S$ a 
$(2n\times 2n)$ symplectic matrix ($\tau S$ is symmetric). The tilda 
symbol denotes vector or matrix transposition. For any operator $O$,
let us denote the average value $\langle O\rangle$ and the shifted
operator $\bar{O}$ by $\langle O\rangle={\rm Tr}\, O{\cal D}/{\cal Z}$
and $\bar{O}=O-\langle O\rangle$. Then, ${\cal D}$ is completely specified
by the knowledge of ``the partition function''
${\cal Z}\equiv\hbox{Tr}\,{\cal D}$, the vector $\am$ and the contraction 
matrix $\rho$ (which we shall also call the single-particle density matrix)
defined as
\be
\label{e6} 
{\rho}_{ij}\equiv\langle(\tau{\bar\al})_j{\bar\al}_i\rangle
.
\ee

For our purpose, the parameterization of the operator $\cal D$ by $\cal Z$,
$\am$ and $\rho$ is more convenient than that by $\nu$, $\la$ and $S$.
In particular, we will see that the evolution equations for the former
variables take a simple form.

With the same notations, we take as our variational choice \cite{BM91,BF93}
for the operator ${\cal A}(t)$ a linear plus quadratic form in the operator
$\al$:
\be
\label{e7}
{\cal A}(t) = \nu_a(t)+\ti{\la}_a(t)\tau\al
+\demi\ti{\al}\tau S_a(t)\al
,
\ee
to which we shall loosely refer as a single-particle operator.

With the choices (\ref{e5}) and (\ref{e7}), it is possible to write
explicitly the functional (\ref{e1}) in terms of the variational
parameters ${\cal Z}$, $\am$, $\rho$, $\nu_a$, $\la_a$ and $S_a$ 
\cite{BM91}. Indeed, we find
\be\label{e77}
\ba{rl}
{\cal I} & ={\rm Tr}\,A{\cal D}(t_1)-\int_{t_0}^{t_1}{\rm d}t \\
&\left(
{\rm Tr}\,{\cal AD}{\ds\dot{{\cal Z}}\over\ds {\cal Z}}
-i{\cal Z}
\left\{
\ti{L}_a\tau (i\dot{\am}-\tau{\ds\partial\ener\over\ds\partial\am})
-\demi {\rm tr}\,S_a (i\dot{\rho}+2[\rho,\,{\ds\partial\ener\over
\ds\partial\rho}])
\right\}
\right),
\ea
\ee
with $L_a=\la_a-S_a\am$. The TDHB equations
\be\label{e8} 
\left\{\ba{rl}
i\dot{{\cal Z}}=& 0 ,\\
i\dot{\am}=& \tau{\ds \partial\ener\over\ds\partial\am} ,\\
i\dot{\rho}=& -2\,\left [\rho,\,{\ds \partial\ener\over\ds\partial\rho}
\right ],
\ea\right.
\ee
are then obtained\cite{BM91,BF93} as the stationarity conditions of the 
reduced form (\ref{e77}) of the BV functional (\ref{e1}) with respect to
the parameters $\nu_a$, $\la_a$ and $S_a$. In these equations,
$\ener\equiv\langle H\rangle$ is the mean-field energy and the dots denote
${\ds {\rm d}\over\ds {\rm d}t}$. We have used the usual convention
$(\partial /\partial\rho)_{ij}=\partial /\partial\rho_{ji}$.

Thus, within the framework of the BV variational principle (\ref{e1}),
and within the trial class (\ref{e5}) for the density operator, the
time-dependent mean-field equations (\ref{e8}) appear to optimize the
expectation values of single-particle observables.

The mean-field energy is a sum of contributions of the type $\rho$ and
$\am\am$ for its kinetic part, $\rho\rho$, $\rho\am\am$ and $\am\am\am\am$
for its potential part. Hence the equations (\ref{e8}) couple the dynamical
variables $\am$ and $\rho$. If this coupling is disregarded, and if the
interaction has zero range, the equation for $\am$ reduces to the
Gross-Pitaevskii equation\cite{gropi}.

The equations (\ref{e8}) do not depend on the parameters $\nu_a (t)$,
$\la_a (t)$ and $S_a (t)$: the evolution of the operator ${\cal D}(t)$
decouples from the evolution of ${\cal A}(t)$. The knowledge of $\nu_a (t)$,
$\la_a (t)$ and $S_a (t)$ is unnecessary for the determination of ${\cal D}(t)$,
and therefore of $\hbox{Tr}A{\cal D}(t_1)$. The solution of eqs.(\ref{e8}) is
uniquely defined once the initial conditions ${\cal Z} (t_0)$, $\am (t_0)$ and
$\rho (t_0)$ are specified. It should be noticed that, in the present context,
this feature results from the linearity in the variational parameters of the
Ansatz (\ref{e7}). Otherwise, in the case of a non-linear parametrization, one
should have to consider simultaneously the evolution equations for ${\cal A}(t)$
obtained by writing the stationarity conditions of (\ref{e1}) with respect to
the parameters characterizing ${\cal D}(t)$. Then, the motions of ${\cal D}(t)$
and ${\cal A}(t)$ would be coupled; moreover, the boundary conditions on
${\cal D}(t_0)$ and ${\cal A}(t_1)$ would lead to a mixed boundary value
problem\cite{BV81,BF93}.

Although we shall not discuss them in the present paper, the equations of
motion of ${\cal A}(t)$ have their own interest; indeed, they are tightly
related to the random phase approximation (RPA) and they play a crucial role
\cite{BV93} for the coherent determination of fluctuations and (two-time)
correlation functions of the single-particle observables (\ref{e7}).

Let us now discuss briefly some general propreties of the TDHB equations
(\ref{e8}) (see \cite{BV85,BM91} for a more detailed discussion).
We note first that the ''partition function'' ${\cal Z}$ is conserved;
this is automatically entailed by the BV variational principle when
the allowed variations $\delta {\cal A}(t)$ includes the unit operator.
Other properties of the TDHB equations are the conservation of the mean-field
energy $\ener$ (for a time-independent Hamiltonian) and the unitary
evolution of the contraction matrix $\rho (t)$. The latter means that the 
eigenvalues of $\rho (t)$ are conserved. This implies in particular the
conservation of the single-particle von-Neumann entropy
${\cal S}=-\hbox{Tr}{\cal D}\log{{\cal D}}$ and that an initially pure state
$\rho\, (\rho +1)=0$ remains pure during the TDHB evolution.
The conservation of the free energy $\Om =\ener - \beta^{-1}{\cal S}$ naturally
follows ($\beta^{-1}$ is the temperature.) We will see the interest of this
property in the next sections.
These conservation laws show that the TDHB approximation reproduces,
in the reduced single-particle space, some of the exact ones.

\disregard{
\be
\label{e10} 
\left\{\ba{rl}
{\cal Z}^{\star} = & {\cal Z} ,\\
\am^{\star} = & \sigma_1\,\am ,\\
\rho^{\dag} = & \sigma_3\,\rho\,\sigma_3 ,
\ea\right.
\ee
where $\sigma_1$ and $\sigma_3$ are the $(2n\times 2n)$ first and 
third Pauli matrices. Equations (\ref{e8}) show that these properties 
are preserved during the evolution if they were initially satisfied.
It is to be mentioned however, that the hermiticity of the operator 
${\cal D}(t)$ is not a necessary requirement of the above derivation
from the expression (\ref{e1}). This freedom might be useful for 
tunneling problems.
The only requirement is that, for a hermitian operator ${\cal A}(t)$,
the expectation value $\hbox{Tr}{\cal AD}(t_1)$, the sole measurable 
quantity, must be real.
}

\setcounter{equation}{0}
\bc{\section{The Lie-Poisson Structure of the TDHB Equations}}\ec

We now examine the mathematical structure of the TDHB equations 
(\ref{e8}). A first interest of this analysis is to establish a close
connection between the TDHB equations and dynamical systems
\cite{marsden,saraceno,KK76}. It may prove useful in the elaboration
of more sophisticated approximations.

It was pointed out in ref.\cite{BV89} that the TDHF equations possess 
a Lie-Poisson structure. The reason is that they deal with the expectation
values of the one-body operators $a_i^{\dag} a_j$ which form a Lie algebra.
In the boson case, the set of the operators $\{1$, $\al$, $\al\al\}$ form
also a Lie algebra. One expects, therefore, the existence of a Lie-Poisson
bracket associated with the TDHB equations (\ref{e8}). This existence results
also directly from a theorem which can be found in refs.\cite{muku,weinstein}).
In order to construct the corresponding Lie-Poisson bracket, let us introduce
the set of the relevant operators $\op^{\mm}$ ($\mm =1,2,3$) defined as follows:
\be\label{eq1}
\op^{1} = 1 \quad ,\quad \op_i^{2} = \al_i \quad ,\quad
\op_{ij}^{3} = (\tau{\bar\al})_j {\bar\al}_i \, .
\ee
These operators satisfy the Lie-algebra
\be\label{eq2}
\left[ \op^{\mm} , \op^{\nn} \right] = \sco_{\sigma}^{\mm\nn} \op^{\sigma}
.\ee
Using the commutation relations (\ref{e4}), we find that the nonvanishing
structure constants $\sco_{\sigma}^{\mm\nn}$ are:
\be\label{eq3}
\left\{
\ba{rl}
(\sco_1^{22})_{ij} = & \tau_{ij} ,\\
(\sco_1^{23})_{i,jk} = & (\sco_1^{32})_{ji,k} = \delta_{ik}\am_j 
-\tau_{ij} (\tau\am )_k ,\\
(\sco_2^{23})_{i,jk}^l = & (\sco_2^{32})_{ji,l}^k = -\delta_{ik}\delta_{jl}
+\tau_{ij} \tau_{kl} ,\\
(\sco_3^{33})_{ij,kl}^{mn} = & \delta_{il} \delta_{jm} \delta_{kn} -
\delta_{jk} \delta_{lm} \delta_{in} +
\tau_{ik} \delta_{jm} \tau_{ln} - 
\tau_{jl} \delta_{in} \tau_{km} 
. 
\ea
\right.
\ee

The Lie-Poisson bracket associated with the TDHB equations can 
now be constructed as follows. Consider the quantities 
$\zoz^{\mm} = \langle\op^{\mm}\rangle$ which are directly related
to our dynamical variables $\cal Z$, $\am$ and
$\rho$. The quantity $\zoz = (\zoz^{1},\zoz^{2},\zoz^{3})$ is a 
vector in the Liouville space. This suggests defining the Poisson 
tensor $\left\{\zoz^{\mm} , \zoz^{\nn}\right\}$ as
\be\label{eq4}
{\cal C}^{\mm\nn} \equiv \left\{\zoz^{\mm} , \zoz^{\nn} \right\} 
= \sco_{\sigma}^{\mm\nn} \zoz^{\sigma}
.\ee
Upon using (\ref{eq3}), we can see that ${\cal C}^{\mm\nn}$ has only two
non-zero components, namely
\be\label{eq5}
\left\{
\ba{rl}
({\cal C}^{22})_{ij} = & \tau_{ij} ,\\
({\cal C}^{33})_{ij,kl} = & \delta_{il} \rho_{kj} -\delta_{kj} \rho_{il}
+\tau_{ik} (\tau\rho)_{lj} - \tau_{lj} (\rho\tau)_{ik} 
. 
\ea
\right.
\ee
Obviously, ${\cal C}^{\mm\nn}$ is antisymmetric with respect to its lower 
indices owing to the antisymmetry of the matrix $\tau$ and to the property 
$\rho_{ji} = (\tau (1+\rho)\tau )_{ij}$ (see \ref{e6}).

We define the Lie-Poisson bracket of two functions $h(\zoz )$ and $g(\zoz )$,
depending on the components of the variables $\zoz^{\mm}$), in the usual
way\cite{muku}:
\be\label{eq6}
\left\{ h , g \right\} = {\ds 1\over\ds i}\, 
{\ds\partial h\over\ds\partial\zoz^{\mm}} {\cal C}^{\mm\nn}
{\ds\partial g\over\ds\partial\zoz^{\nn}}
.
\ee
This expression is clearly bilinear and antisymmetric. In addition, it 
satisfies Leibniz's derivation rule
\be\label{eq7}
\left\{ hg , k \right\} = h\left\{ g , k \right\}+\left\{ h , k \right\}g
,
\ee
and Jacobi's identity. This last property follows directly from the Lie
algebra (\ref{eq2}). Replacing in (\ref{eq6}) the function $h$ by one component
of the variables $\zoz$ and using (\ref{eq5}), we obtain for any function $g$
\be\label{eq8}
\left\{
\ba{rl}
\left\{ 1 , g \right\} = & 0 ,\\
\left\{ \am_i , g \right\} = & {\ds 1\over\ds i}
(\tau {\ds\partial g\over\ds\partial\am})_i ,\\
\left\{ \rho_{ij} , g \right\} = & -{\ds 2\over\ds i}
[ \rho , {\ds\partial g\over\ds\partial\rho}]_{ij}
. 
\ea
\right.
\ee
In particular, when $g$ is the energy or the free energy, we can express
the TDHB equations in the form
\be\label{eq9}
\dot{\zoz}^{\mm} = \left\{ \zoz^{\mm} , \ener (\zoz )\right\}
= \left\{ \zoz^{\mm} , \Om (\zoz )\right\}
,
\ee
which can also be written as
\be\label{eq10}
\dot{\zoz}^{\mm} = {\ds 1\over\ds i}\, 
{\cal C}^{\mm\nn} {\ds\partial\Om \over\ds\partial\zoz^{\nn}}
.
\ee
The expressions (\ref{eq9}$-$\ref{eq10}) clearly exhibit the classical 
structure of the TDHB equations, with the mean-field energy $\ener$, or
the free energy $\Om$, playing the role of an Hamiltonian. However, the
non-symplectic Poisson tensor (\ref{eq5}) induces a non-canonical
hamiltonian structure (also called a Lie-Poisson system).

From (\ref{eq10}), one sees that the evolution of any differentiable 
function $h(\zoz )$ is governed by the equation
\be\label{eq11}
{\ds {\rm d} h(\zoz )\over\ds {\rm d}t} = \left\{ h(\zoz ) , \Om (\zoz )
\right\}
.
\ee
In particular, any function $h(\zoz )$ which ''Poisson-commutes'' with
the free energy (or the energy) is a constant of the TDHB motion.

\setcounter{equation}{0}
\bc{\section{Linearized Equations and Stability Analysis}}\ec

Let $\zoz_0$ be a static solution of the TDHB equations (\ref{eq9}):
\be\label{eq12}
\left\{ \zoz_0^{\mm} , \Om (\zoz_0 )\right\} = {\ds 1\over\ds i}\, 
{\cal C}^{\mm\nn}(\zoz_0) {\ds\partial\Om \over\ds\partial\zoz_0^{\nn}}
= 0
.
\ee
We consider a small amplitude motion $\dz$ around $\zoz_0$ such that
$\zoz^{\mm} = \zoz_0^{\mm} +\dz^{\mm}$. Using (\ref{eq9}) and
retaining only linear terms in $\dz$, one obtains
\be\label{eq13}
{\ds {\rm d}\over\ds {\rm d}t}\,\left(\dz^{\mm}\right)={\ds 1\over\ds i}\,
{\cal C}^{\mm\nn}(\zoz_0) {\ds\partial\Om_2 \over\ds\partial (\dz^{\nn})}
,
\ee
where $\Om_2$ is the expansion up to second order of the free energy $\Om$:
\be\label{eq14}
\Om_2 (\zoz )=\Om (\zoz_0 ) +
{\ds\partial\Om\over\ds\partial\zoz_0^{\mm}}\dz^{\mm} + \demi
{\ds\partial^2\Om\over\ds\partial\zoz_0^{\mm}\partial\zoz_0^{\nn}}
\dz^{\nn}\dz^{\mm}
.
\ee
The linearized TDHB equations (LTDHB) can therefore be written 
in the form
\be\label{eq15}
{\ds {\rm d}\over\ds {\rm d}t}\,\left(\dz^{\mm}\right)=
\left\{
\dz^{\mm} , \Om_2 (\zoz )
\right\}_0
,
\ee
the Lie-Poisson bracket $\{,\}_0$ being given by (\ref{eq6}) with the 
Poisson tensor evaluated at the point $\zoz_0$. The LTDHB equations
also exhibit a classical structure with $\Om_2$ playing the role of an
Hamiltonian (which we note is a constant of the LTDHB motion) and with
the non-symplectic Poisson tensor ${\cal C}^{\mm\nn}(\zoz_0)$.

The stability analysis of TDHB and LTDHB relies on the Lagrange-Dirichlet
theorem (ref.\cite{BV89} sect. 3.2). The TDHB motion is Lyapunov stable (that
is, a trajectory starting in the neighborhood of $\zoz_0$ remains in this
neighborhood during the evolution) if $\zoz_0$ is a local minimum of $\Om$.
The same condition (with $\Om$ replaced by $\Om_2$) ensures the Lyapunov
stability of the LTDHB motion provided the stability matrix
${\cal B}_{\mm\nn} ={\ds\partial^2\Om\over\ds\partial
\zoz^{\mm}\partial\zoz^{\nn}}\Big|_{\zoz=\zoz_0}$ has only
positive eigenvalues. In this case, the TDHB equations are said to be 
linearly stable. One notes that Lyapunov stability and linear stability are
independent of each other since the non-linear terms in the TDHB equations
may have stabilizing or destabilizing effects on the dynamics.

Let us finally mention that the previous results hold both for pure states
and at non-zero temperature. In the first case, one should evidently replace
the Helmholtz free energy $\Om$ by the mean-field energy $\ener$.

\setcounter{equation}{0}
\bc{\section{Search for Conjugate Variables}}\ec

We have seen in sections 3 and 4 that both TDHB and LTDHB admit a 
non-canonical hamiltonian structure. One may therefore wonder about 
the existence of canonically conjugate variables (CCV). This analysis
is of course of great interest in the study of the classical limit or
in the ''re-quantization'' process of the mean-field approximation.

In principle, the CCV are found by diagonalizing the Poisson tensor
(\ref{eq4}). The second block ${\cal C}^{22}$ being $\zoz$-independent
(see \ref{eq5}), one can easily find a global canonical parametrization
for $\zoz^2$. Indeed, by using the usual cano\-nical trans\-formations 
(where $j=1 \ldots n$)
\be\label{e11}
a_{j}=\rac \, \left ( x_{j}+i\, p_{j} \right )
\quad ,\quad
a_{j}^{\dag}=\rac \, \left ( x_{j}-i\, p_{j} \right )
,
\ee
(which underlie the equivalence between the algebra of boson operators
and the symplectic algebra of canonical pairs of the position and momentum
operators $x$ and $p$), and upon defining the variables
\be\label{e14}
\left\{\ba{rl}
\ffi_i & = \langle x_i \rangle , \\
\py_i & = \langle p_i \rangle,    
\ea\right.
\ee
the evolution of $\am$ translates into the equations
\be\label{e15} 
\left\{\ba{rl}
\dot{\ffi} & = {\ds\partial\Om\over\ds\partial \py}, \\
\dot{\py} & = -{\ds\partial\Om\over\ds\partial \ffi},
\ea\right.
\ee
which are clearly in an Hamiltonian form, and which, we recall, are coupled
to $\rho$ through $\Om$.

In order to discuss the possibility of a canonical parametrization of
$\zoz^3$, one must note that the block ${\cal C}^{33}$ depends on the 
point $\zoz$ (more precisely, on $\rho$). One looks therefore for {\it local}
CCV (there is no rigorous proof of non-existence, but the search for
{\it global} sets of CCV seems very hazardous.) Let us consider
the instructive (and simpler) example of the LTDHB equations (\ref{eq15})
in a basis where $\rho_0$ is diagonal: 
$\rho_{0ij} =\rho_{0i}\delta_{ij}$. The equation (\ref{eq5}) gives
\be\label{eq16}
\left( {\cal C}^{33}\right)_{ij,kl} = (\rho_{0j}-\rho_{0i})
\left(\delta_{il}\delta_{kj}+\tau_{ik}\tau_{lj}\right),
\ee
and the LTDHB equations write
\be\label{eq17}
{\ds {\rm d}\over\ds {\rm d}t}\delta\rho_{ij}=i (\rho_{0i}-\rho_{0j})
\left(
{\ds\partial\Om_2 \over\ds\partial (\delta\rho)} +\tau
{\ds\partial\Om_2 \over\ds\partial (\delta\ti\rho)}\tau
\right)_{ij}
.
\ee
It is easily seen that when $\rho_{0i}-\rho_{0j}>0$, one can define the 
variables
\be\label{eq18}
q_{ij}={\ds\delta\rho_{ij}+\delta\rho_{ji}\over\ds
\sqrt{2(\rho_{0i}-\rho_{0j})}}
\quad,\quad
p_{ij}={\ds i(\delta\rho_{ij}-\delta\rho_{ji})\over\ds
\sqrt{2(\rho_{0i}-\rho_{0j})}},
\ee
of which the evolution takes the form of Hamilton's equations
\be\label{eq19}
{\dot q}_{ij} = {\ds\partial\Om_2 \over\ds\partial p_{ij} } 
\quad,\quad
{\dot p}_{ij}=-{\ds\partial\Om_2 \over\ds\partial q_{ij}} .
\ee
However, for $\rho_{0i}=\rho_{0j}$ (degeneracy in $\rho$), we see on 
(\ref{eq17}) that one must introduce new variables $r_{ij}=\delta\rho_{ij}$
which are associated to the eigenvalue 0 of the Poisson tensor.

This scheme is general. When the Poisson tensor possesses vanishing
eigenvalues (related to the existence of constants of the motion),
it is unavoidable, in order to obtain a canonical parametrization,
to introduce additional variables that Poisson-commute with the CCV.
Let us illustrate this point by looking for a canonical parametrization
of $\rho$ in the pure state case ($\beta^{-1}=0$). Using (\ref{e11}),
the matrix $\rho$ can be expressed in terms of the ($n\times n$) matrices
(see \cite{BM91})
\be\label{eq20}
\left\{
\ba{rl}
\Ga_{ij} & = \xij , \\
\Lam_{ij} & = \pij ,\\
\De_{ij} & = \xipj ,
\ea\right.
\ee
of which the evolution equations write
\be
\label{e16} 
\left\{\ba{rl}
\dot{\Ga} & = 2\,(2
{\ds\partial\Om\over\ds\partial\De}\Ga + 
\De {\ds\partial\Om\over\ds\partial\Lam}
)_{s}, \\
\dot{\Lam} & =-2\,(2
\Lam {\ds\partial\Om\over\ds\partial\De} + 
{\ds\partial\Om\over\ds\partial\Ga}\De
)_{s}, \\
\dot{\De} & =-2\,[\De ,{\ds\partial\Om\over\ds\partial\De}] 
-4\,
(
\Ga {\ds\partial\Om\over\ds\partial\Ga}
-{\ds\partial\Om\over\ds\partial\Lam}\Lam
) 
,
\ea\right.
\ee
where the index $s$ denotes the symmetric parts of the matrices involved.
The pure state property $\rho (\rho +1)=0$ is equivalent to the conditions
\be\label{e17} 
\left\{\ba{rl}
I_{ij} & \equiv 4\,(\Ga\Lam )_{ij} -(\De^{2})_{ij} = \delta_{ij},\\
(\De\Ga )_{ji} & = (\De\Ga )_{ij},\\
(\Lam\De )_{ji} & = (\Lam\De )_{ij} .
\ea\right.
\ee
As already mentioned in section 2, these conditions are preserved by the 
equations (\ref{e16}) if they were initially fulfilled. This is a typical
example of vanishing eigenvalues of the Poisson tensor which depend on the 
initial conditions. Having chosen these initial conditions, one may in 
principle find a canonical form for the equations (\ref{e16}) with the 
additional constraints (\ref{e17}). But in our particular case, we can 
explicitly eliminate these constraints by defining the two symmetric 
matrices
\be\label{e21} 
\left\{\ba{rl}
\FF & = \Ga ,\\
\GG & = {\ds 1\over\ds 4}\,\Ga^{-1}\De .
\ea\right.
\ee
Indeed, the evolution of $\FF$ and $\GG$ is governed by Hamilton's 
equations
\be\label{e22} 
\left\{\ba{rl}
\dot{\FF} & = {\ds\partial\ener\over\ds\partial \GG}, \\
\dot{\GG} & =-{\ds\partial\ener\over\ds\partial \FF} ,
\ea\right.
\ee
and $\Lam$ will be given by 
$\Lam = {\ds 1\over\ds 4}\,\FF^{-1} + 4\, \GG\FF\GG$. It should be noticed
however, that this canonical parametrization is not global since it does not
span the whole manifold $\rho$, but is rather valid on a {\it sheet}
characterized by the special initial conditions.

Our last illustration in the search for CCV is the one-dimensional case.
The matrices defined in (\ref{eq20}) become c-numbers.
One can see that the quantity $I$ is conserved, whatever the initial
conditions. In particular, the initial state needs not be pure. In order
to rewrite the TDHB equations, it is possible to keep the definitions
(\ref{e21}). Equations (\ref{e22}) then remain valid (with $\Lam$ replaced by
$I/4\FF +4 \FF\GG^2$). But sometimes, it is preferable to work with 
variables that have the dimensionalities of $x$ and $p$. Hence, if we define 
\be\label{e23}
\left\{\ba{rl}
\ww & = \sqrt{\Ga}, \\
\uu & = \demi {\ds \De\over\ds\sqrt{\Ga}},
\ea\right.
\ee
we obtain from (\ref{e16})
\be\label{e24}
\left\{\ba{rl} 
\dot{\ww} & = {\ds\partial\Om\over\ds\partial \uu}, \\
\dot{\uu} & =-{\ds\partial\Om\over\ds\partial \ww}.
\ea\right.
\ee
We arrive at the conclusion that a quantum one-dimensional bosonic system 
in the TDHB ap\-proximation evolves as a classical Hamiltonian system with
two coupled degrees of freedom. In contrast with the $n-$dimensional case,
this result holds for systems at non-zero temperature.

\setcounter{equation}{0}
\bc{\section{Conclusions}}\ec

Upon selecting a trial time-dependent density operator ${\cal D}(t)$ 
belonging to the class of gaussian bosonic operators and a trial 
time-dependent observable ${\cal A}(t)$ belonging to the class of
''one-body'' operators, we derived, from the BV variational principle,
the evolution equations (labeled by TDHB) for the first three 
contractions of ${\cal D}(t)$ namely, the ''partition function'', 
the expectation value $\am$ of the ''one-boson'' operator $\al$ and the 
contraction matrix $\rho$. 

We have investigated the relation of these TDHB equations with the 
hamitlonian formalism. We have exhibited the Lie-Poisson bracket 
associated with the TDHB equations and with their linearized form (LTDHB) 
by exploiting the Lie algebra of the set $\{1$, $\al$, $\al\al\}$ of the 
relevant observables. We have also performed a stability analysis of TDHB 
and LTDHB.

The fact that the Poisson tensor associated with this bracket
is non-symplectic, brings forward the main difference between the TDHB 
and Hamiltonian evolutions, namely, that the former is non-canonical.
This led us to wonder about the possible existence of canonically conjugate 
variables. The very existence of vanishing eigenvalues of the Poisson tensor,
the dependence of the latter on the point of the manifold and, last but not
least, the variation of the multiplicity of the zero eigenvalues from point to
point, render the search for global sets of CCV highly non-trivial in the
general case. Local canonical parametrizations are however possible in spite of the
additional constraints that one must introduce to account for the conservation
equations. This was shown to be the case for the LTDHB equations. We have
also shown in this context, that the TDHB equations, in the pure state limit,
can be cast in an Hamiltonian form by removing the constraints associated
with the pure state conditions, and therefore reducing the number of degrees 
of freedom.

The study of the mathematical properties of the TDHB approximation may also
prove useful in the search for consistent extensions. For instance, one could
''re-quantize'' the TDHB equations by suitably deforming their associated
Lie-Poisson bracket.

We are indebted to R. Balian, M. V\'en\'eroni, C. Martin, H. Flocard and D.
Vautherin for fruitful discussions. 

\end{document}